\documentclass[%
reprint,
frontmatterverbose, 
nofootinbib,
amsmath,amssymb,
aps,showkeys,showpacs,
prd,
prstab,
prstper,
]{revtex4-1}
\usepackage{newtxmath,newtxtext}
\usepackage{graphicx}
\usepackage[caption=false]{subfig}
\usepackage{dcolumn}
\usepackage{bm}
\usepackage{dsfont}
\usepackage{amsmath,amssymb,stackengine}
\usepackage{siunitx}
\usepackage[caption=false]{subfig}
\usepackage{braket}
\usepackage{listings}
\usepackage{xcolor}
\usepackage{hyperref}
\usepackage{tcolorbox}
\usepackage{braket}
\hypersetup{colorlinks=true,linkcolor=blue,urlcolor=blue,citecolor=blue}
\usepackage{graphicx}% Include figure files
\usepackage{dcolumn}% Align table columns on decimal point
\usepackage{bm}% bold math
\usepackage{orcidlink}
\usepackage{multirow, makecell}
\usepackage{rotating}

\begin{document}
\title{Entropy Bounds and Holographic Dark Energy: Conflicts and Consensus}

\author{Manosh T. Manoharan\orcidlink{0000-0001-8455-6951}}
\email{tm.manosh@gmail.com; tm.manosh@cusat.ac.in}
\affiliation{Department of Physics, Cochin University of Science and Technology, Kochi -- 682022, India.}%


\begin{abstract}
	Cohen, Kaplan, and Nelson's influential paper established that the UV-IR cut-offs cannot be arbitrarily chosen but are constrained by the relation $\Lambda^2 L \lesssim M_p$. Here, we revisit the formulation of the CKN entropy bound and compare it with other bounds. The specific characteristics of each bound are shown to depend on the underlying scaling of entropy. Notably, employing a non-extensive scaling with the von Neumann entropy definition leads to a more stringent constraint, $S_{\text{max}} \approx \sqrt{S_{\text{BH}}}$. We also clarify distinctions between the IR cut-offs used in these frameworks. Moving to the causal entropy bound, we demonstrate that it categorises the CKN bound as matter-like, the von Neumann bound as radiation-like, and the Bekenstein bound as black hole-like systems when saturated. Emphasising cosmological implications, we confirm the consistency between the bounds and the first laws of horizon thermodynamics. We then analyse the shortcomings in standard Holographic Dark Energy (HDE) models, highlighting the challenges in constructing HDE using $\Lambda^2 L \lesssim M_p$. Specifically, using the Hubble function in HDE definitions introduces circular logic, causing dark energy to mimic the second dominant component rather than behaving as matter. We further illustrate that the potential for other IR cut-offs, like the future event horizon in an FLRW background or those involving derivatives of the Hubble function, to explain late-time acceleration stems from an integration constant that cannot be trivially set to zero. In brief, the CKN relation doesn't assign an arbitrary cosmological constant; it explains why its value is small. 
\end{abstract}

\keywords{Cosmological Constant Problem, Entropy Bounds, Holographic Dark Energy, Horizon Thermodynamics}
\maketitle

\section{Introduction}

The cosmological constant (CC) remains the simplest and most favoured candidate for dark energy, explaining the observed late-time cosmic acceleration \cite{RevModPhys.75.559}, whose most plausible physical origin is the vacuum energy density. However, quantum field theory (QFT) calculations require an unnatural degree of fine-tuning—of the order of $ 10^{120} $—if QFT is assumed valid up to the Planck scale. Even relaxing this assumption leaves several orders of magnitude of fine-tuning unresolved. This profound discrepancy is known as the old cosmological constant problem \cite{RevModPhys.61.1}.  

Additionally, the constancy and dominance of the vacuum energy density in the present epoch raise further questions. Specifically, why is the matter density comparable to the CC density today? This constitutes the cosmic coincidence problem. These challenges, collectively termed the cosmological constant problem, have remained unresolved. Resolving them requires a mechanism that can predict the value of the CC from a QFT or effective QFT framework while explaining its dominance in the current epoch \cite{PhysRevLett.82.896}.  

Dynamic dark energy models address the coincidence problem by introducing a time-dependent energy density that evolves to dominate at late times \cite{doi:10.1142/S0218271812300029}. However, this approach merely shifts the fine-tuning problem to the initial conditions of the dynamical field, leaving the more profound issue—a QFT origin of the CC—unresolved. Hence, any dynamic dark energy model cannot claim prominence over the CC without stating its origin \cite{Wang_2016}.  

In \cite{PhysRevLett.82.4971}, the authors incorporate the ideas of the holographic principle \cite{RevModPhys.74.825} and claim that any observed vacuum energy density must be smaller than the energy density of a black hole of the same size. Thus connecting vacuum energy density with an entropy bound. Translating this notion of size to an infrared (IR) cut-off yields an energy density consistent with observations when the IR cut-off is set to $ 1/H_0 $, where $ H_0 $ is the Hubble function at $z=0$. 

Subsequent critiques in \cite{HSU200413} and \cite{LI20041} highlighted issues with this approach. These works argued that using $ 1/H $ as the IR cut-off results in an incorrect equation of state, necessitating alternative cut-offs (note that instead of $H_0$ as in \cite{PhysRevLett.82.4971}, authors of \cite{HSU200413} used $H$). These discussions laid the basis for holographic dark energy models, which extend the holographic principle to construct evolving dark energy scenarios \cite{PAVON2005206}.  

With an increasing trend to resolve various cosmological tensions, most dark energy models often overlook why and how such a model was established. This article aims to critically study the constructions of HDE models, particularly their capacity to address the original CC problem.

The vacuum energy, entropy bounds, and cosmological constant are intricately connected, contributing to the development of holographic dark energy models. However, drawing unnecessary connections between these concepts can result in misleading deductions. This article aims to clarify existing misunderstandings and provide logical explanations for the origin and formulation of entropy bounds and holographic dark energies. This article is divided into two major sections. In the first half, we will discuss the features and applications of various entropy bounds, highlighting their differences and appropriate contexts for use. In the second half, we will examine several prominent holographic dark energy models, identifying significant concerns and logical inconsistencies in their construction and application. 

\section{Entropy Bounds -- Origin and Assumptions}

In standard thermodynamics and statistical mechanics, entropy has a well-defined meaning in terms of the number of possible states. This definition enables the calculation of a system's entropy using either thermodynamic or statistical principles, which are equivalent under the Boltzmann distribution. However, these methods are limited to systems without long-range interactions, such as gravity. 

The most well-known bound on a system's entropy was identified by Bekenstein \cite{PhysRevD.23.287}, building on earlier works on black hole entropy \cite{PhysRevD.7.2333} and the generalised second law \cite{PhysRevD.9.3292}. For a system with radius $R$ and energy $E$, this bound constrains the system's entropy as \cite{PhysRevD.23.287},
\begin{equation}
S\leq 2\pi ER/\hbar, 
\label{eq:BekensteinBound}
\end{equation}
Interestingly, as it does not involve Newton's constant, it applies to systems with negligible or no gravity while reaching saturation for highly gravitating objects like black holes. This bound arises naturally from the generalised second law, a concept further explained by Bousso's more general covariant entropy conjecture \cite{Raphael_Bousso_1999}. It is important to note that $S$, $E$, and $R$ pertain to the same system. The right-hand side of the bound should not be mistaken for black hole entropy unless the left-hand side represents the entropy of a black hole, in which case the bound is saturated.

Switching this argument, one can, in principle, establish a bound on the energy of a system of size $R$ at saturation, knowing that the system's entropy will match the entropy of a black hole of the same size. This provides a bound on the energy and, consequently, the energy density, which addresses the issue of ultraviolet divergence. This crucial insight is the core rationale behind the celebrated work by Cohen, Kaplan, and Nelson (hereafter CKN) \cite{PhysRevLett.82.4971}.

The conventional flat space quantum field theory (QFT) estimate of zero-point energy ($\Delta E_{vac}$) exhibits both ultraviolet (UV) and infrared (IR) divergences. This arises primarily from the assumption that QFT is valid up to arbitrarily large energy and volume. In \cite{PhysRevLett.82.4971}, CKN demonstrate that even if the UV cut-off is chosen to be as low as possible, an infinite volume allows the system to collapse into a black hole. They showed that a sensible (effective) QFT can only be constructed within a finite volume with a UV-IR connection instead of local constraints. The effective QFT is then defined up to the IR cut-off, which also serves as the system's boundary. What lies beyond the IR cut-off is excluded in CKN's description. Thus, they established that,
\begin{equation}
\Lambda\lesssim\sqrt{\frac{M_p}{L_{IR}}},
\end{equation}
where $\Lambda$ is the UV energy cut-off, $M_p$ is the Planck mass, and $L_{IR}$ is the IR length cut-off. One could also work with the UV length cut-off $\epsilon \sim 1/\Lambda$ as a minimum lattice parameter.\\

\textit{How is this different from the standard results?} \\

In conventional QFT, for a scalar field, we have,
$\rho_{vac}=\Delta E_{vac}/V(\mathbb{R})\propto\int_{0}^{\Lambda}p^2dp\sqrt{p^2+m}$. We get $\rho_{vac} \simeq \Lambda^4$ for a massless scalar field. The problem with this definition is that it does not properly account for the volume $V(\mathbb{R})$, which can, in principle, be infinite. The CKN bound breaks this image by invoking the maximum energy in a given volume, thereby addressing the IR divergence and solving the UV-IR divergence concurrently. In doing so, we find that $\Lambda^4 \lesssim {M_pL_{IR}}/{L_{IR}^3}$. 

Using the current Hubble horizon as the IR cut-off, CKN relation yields $\Lambda^4 \lesssim (10^{-2.5} \text{eV})^4$, producing a result comparable with current observations and thus eliminating the need for unnatural fine-tuning. This, however, is not a prediction, as there is no explicit reason to choose the Hubble horizon as the IR cut-off, and there is always the freedom to add an arbitrary constant. Physically, the bound implies that the energy cannot surpass $M_p / L^2$, with any value below being feasible. Probabilistically, zero appears to be the most favoured value \cite{PhysRevLett.85.1610}. The idea is to avoid all possible black hole states and accept that the entropy is \textit{extensive}.

\subsection{Cosmologists versus CKN}

A cosmologist starting from the Friedmann equation may argue that this observation can be derived from the first Friedmann equation. Precisely, a cosmologist would estimate $\rho_{vac} \lesssim (10^{-2.5} \text{eV})^4$ (as Weinberg \cite{RevModPhys.61.1}), noting that this is significantly lower than standard QFT predictions. \textit{It is then tempting to argue that the CKN relation and the first Friedmann equation are the same}, but that is just wrong.

CKN discovered that the conventional QFT prediction is not well-posed and that a UV-IR regulator is necessary. Here, the horizon entropy acts as the regulator, yielding an estimate compatible with cosmological observations. Thus, the CKN bound provides an effective upper bound on the vacuum energy density within the Hubble volume. Although CKN assumed a fixed Hubble horizon (e.g., $H_0$) rather than a dynamic one ($H$), they did so without presupposing any specific cosmological models. The CKN relation concerns only the vacuum contribution to an effective QFT with a UV cut-off $\Lambda$ and an IR cut-off $L_{IR}$, assuming no strong gravitational sources like black holes. It was developed based on the premise that QFT calculations can be performed without considering gravity.

Let us look at the above concern in a simple de Sitter setting. The de Sitter vacuum energy density using the Friedmann equation is given as,
\begin{equation}
\rho_{dS: vac}=\frac{3}{8\pi G}H^2_{dS},
\end{equation}
where $H^2_{dS}$ is the constant Hubble parameter connected with the cosmological constant. If we consider $H^2_{dS}$ as the IR cut-off, from the CKN relation, we get, 
\begin{equation}
\rho_{vac}\lesssim \frac{3}{8\pi G}H^2_{dS}.
\end{equation}
The way CKN avoids the fine-tuning is such that,
\begin{equation}
\rho_{vac}\lesssim\rho_{dS: vac}
\end{equation}
Thus, given there is a cosmological constant contributing to $\rho_{dS: vac}$, the constant vacuum energy density $\rho_{vac}$ will be close to it such that what we observe is
\begin{equation}
\rho_{eff: vac}=|\rho_{dS: vac}+\rho_{vac}|\sim(10^{-2.5} \text{eV})^4
\end{equation}
And since, $\rho_{vac}\lesssim\rho_{dS: vac}$, there is no fine-tuned cancellation.

The question deepens when we bring the notion of entropy of de Sitter space. The milestone work by Gibbons and Hawking \cite{PhysRevD.15.2752} showed us that the de Sitter space has a boundary term in the action contributing to its horizon entropy, $S_{dS}=\pi/H_{dS}^2$. If we rewrite the CKN relation as $\rho_{vac}\lesssim S/L_{IR}^4$, and identity $S$ with $S_{dS}$ and $L_{IR}$ with $1/H_{dS}$, we will have, $\rho_{vac}\lesssim 1/S_{dS}$. At saturation, this is the connection between de Sitter entropy and cosmological constant \cite{PhysRevD.15.2738}. So, the CKN energy relation is built into the laws of thermodynamics, and the thermodynamic law establishing this connection is the first law. Thus, the CKN relation at saturation is consistent with the first law of horizon thermodynamics. One must understand that, this energy condition is not the CKN entropy bound.

\subsection{IR cut-off vs Boundary}

To understand the distinction between CKN's findings and other entropy bounds, we must first grasp CKN's definition of entropy. In \cite{PhysRevLett.82.4971}, CKN define entropy based on the number of unit cells in a lattice. For a system of size $L_{IR}$ with lattice spacing $1/\Lambda$, the number of unit cells is given by $N=L_{IR}^3/(1/\Lambda^3) = L_{IR}^3\Lambda^3$. Considering there are $N$ cells, and each corresponds to a bit of quantum information (such as a qubit with base two), the entropy can be expressed in units of $\log_b$ as $S = \log_{b}b^N = N$, where $b$ is the base. Therefore, it is natural to assume, as CKN did, that $S \sim L_{IR}^3\Lambda^3$.

To further explore the concept, consider a total system defined by a pure state. By definition, such a system has \textit{zero entropy}. If this system comprises non-interacting subsystems, each subsystem remains pure, maintaining zero entropy. However, if interactions are present (regardless of their range), the subsystems can evolve into non-pure states, creating quantum or classical correlations. This process generates entropy within the subsystems, often called entanglement entropy or quantum discord \cite{RevModPhys.84.1655, RevModPhys.81.865}. Despite these interactions, the total system remains unitary, retaining zero entanglement entropy, as there is nothing external with which it can become entangled. Thus, to define entropy based on these notions, we must introduce an additional element in the description: \textit{the boundary}. Introducing this boundary divides the total system into distinct parts, allowing us to define the entanglement entropy between these parts as a function of the respective boundaries.

When CKN refers to the IR cut-off, they indicate the total box size within which we perform the QFT. Although there can also be a concept of a boundary or horizon that lies below the actual IR cut-off, for CKN, there are no such interpretations. They consider an extensive entropy that scales with the volume, leading to the result that the maximum entropy of the box, $S_{\text{max}}$, is approximately $S_{BH}^{3/4}$, where $S_{BH}$ is the entropy of a black hole of the same size. Therefore, \textit{unlike a pure quantum state, the total entropy of the system in CKN is not zero}.

\begin{figure}[h]
	\centering
	\includegraphics[width=0.4\textwidth]{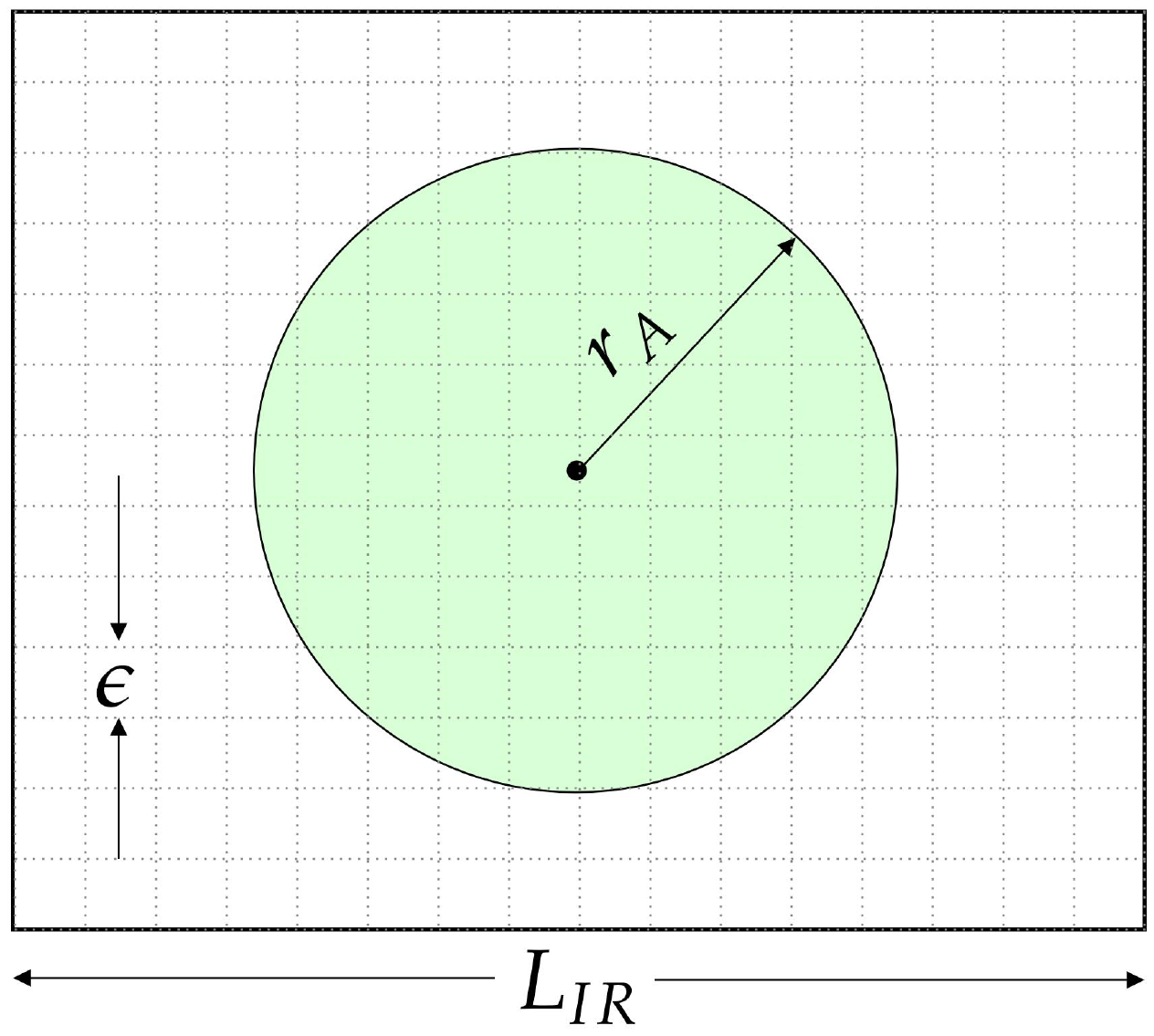}
	\caption{A lattice of IR cut-off $L_{IR}$ and UV cut-off $\epsilon\sim1/\Lambda$ along with an arbitrary boundary at $r_{A}$.\label{fig:UVRAIR}}
\end{figure}

Now, consider a scenario where we have both UV and IR cut-offs (denoted as $\epsilon$ and $L_{IR}$, respectively) along with an arbitrary boundary $r_A$, such that $\epsilon < r_A < L_{IR}$. We then ask, what is the entropy of the region within or outside $r_A$? This question is particularly significant in cosmology because the Hubble horizon is not necessarily the IR cut-off of the universe; the actual universe could be several orders of magnitude larger than what we can observe (or infinite). 

Furthermore, unlike AdS space, de Sitter space does not possess a finite boundary, and there have been limited advancements in developing a statistical framework for horizon entropy in such framework \cite{Banihashemi2022}. Thus, the Hubble horizon (or the apparent horizon in non-flat cases) can be considered a causal length scale, allowing observers to consider it as a thermodynamic system \cite{SANCHEZ2023137778}. Therefore, the Hubble horizon is a boundary which satisfies the laws of thermodynamics, while the exact IR cut-off remains irrelevant (see FIG.\ (\ref{fig:UVRAIR})). 

If we regard the Hubble horizon as the IR cut-off and adopt CKN's entropy definition, the CKN bound gives us the upper limit of the probable energy density. Conversely, when exploring the von Neumann entropy definition, the boundary defines the entropy independently of the IR cut-off. One might initially dismiss this as a mere difference in terminology. However, the von Neumann entropy, defined in terms of the entanglement entropy of a quantum system, imposes a more stringent constraint than CKN.

\section{Von Neumann Entropy Bound}

Let's examine how the choice of entropy definition influences the entropy bound according to the calculations presented by CKN. Regarding the entropy bound, one can express the CKN relation as,
\begin{align*}
\Lambda^4L^3&\lesssim LM_p^2 \text{, (Energy bound)}\\
\implies\Lambda^4L^4&\lesssim L^2M_p^2\\\implies \Lambda^3L^3&\lesssim (L^2M_p^2)^{3/4}\text{, (assumed extensive entropy)}
\end{align*}
Here, $\Lambda^4 L^3$ denotes the mass within the volume $L^3$, and $L M_p^2$ represents the mass of a black hole of the same size. Although it is permissible to assume a length scale larger than $L$, in this scenario, we consider only the energy enclosed within the volume $L^3$ and constrain it by the energy of a black hole of the same size. Thus, unlike the Bekenstein bound, the left and right sides differ.

If we identify $\Lambda^3 L^3$ as the entropy of the system, as CKN did, we will obtain,
\begin{align}
S_{\text{max}} \lesssim S_{\text{Black hole}}^{3/4} \text{, (The CKN Bound)}.
\end{align}
However, in the framework we illustrated earlier, where the UV/IR cut-off is considered along with a boundary, the definition of entropy takes on a different form. Suppose the quantum field is perfectly contained within a box of size $L_{IR}$ by some wavefunction, with no apparent decay outside the box. In that case, the system adheres to unitary dynamics, and the von Neumann entropy of the whole system is zero by definition. Nevertheless, the regions inside and outside a boundary ($r_A$) can each have non-zero entropy, which is, by definition, equal. As the boundary evolves towards its maximum size, the subsystem becomes the entire system, and the external region ceases to exist.
Consequently, the outside and inside entropy become zero, matching the complete system's entropy. This self-consistent definition of entropy enables us to compute the entropy for any arbitrary boundary. One can show that this adheres to an area law as \cite{PhysRevLett.71.666},
\begin{equation}
S_{\text{inside}} = k\Lambda^2r_{A}^2.
\label{eq:vnentroy}
\end{equation}
Here, $\Lambda = 1/\epsilon$ in natural units is the UV cut-off, and $k$ is an order one factor.

Following the energy-bound arguments employed by CKN, we derive the following,
\begin{align*}
\Lambda^4 r_{A}^3 \lesssim r_{A} M_p^2 < L_{IR} M_p^2 \implies \Lambda^4 r_{A}^4 \lesssim r_{A}^2 M_p^2.
\end{align*}
Now, using the von Neumann entropy as in Eq. (\ref{eq:vnentroy}), we find,
\begin{equation}
S_{\text{inside (max)}} \simeq S_{\text{Black hole}}^{1/2}\text{, (The von Neumann Bound).}
\end{equation}
Thus, the maximum entropy within a region of space containing some quantum field (such as a massless scalar field) scales as the square root of the entropy of a black hole of the same size. This is an unexplored observation, imposing a more stringent constraint than the one proposed by CKN. The critical difference between this result and CKN lies in the choice between extensive and non-extensive entropy and the adoption of the von Neumann definition.
\begin{figure}[h]
	\centering
	\includegraphics[width=0.4\textwidth]{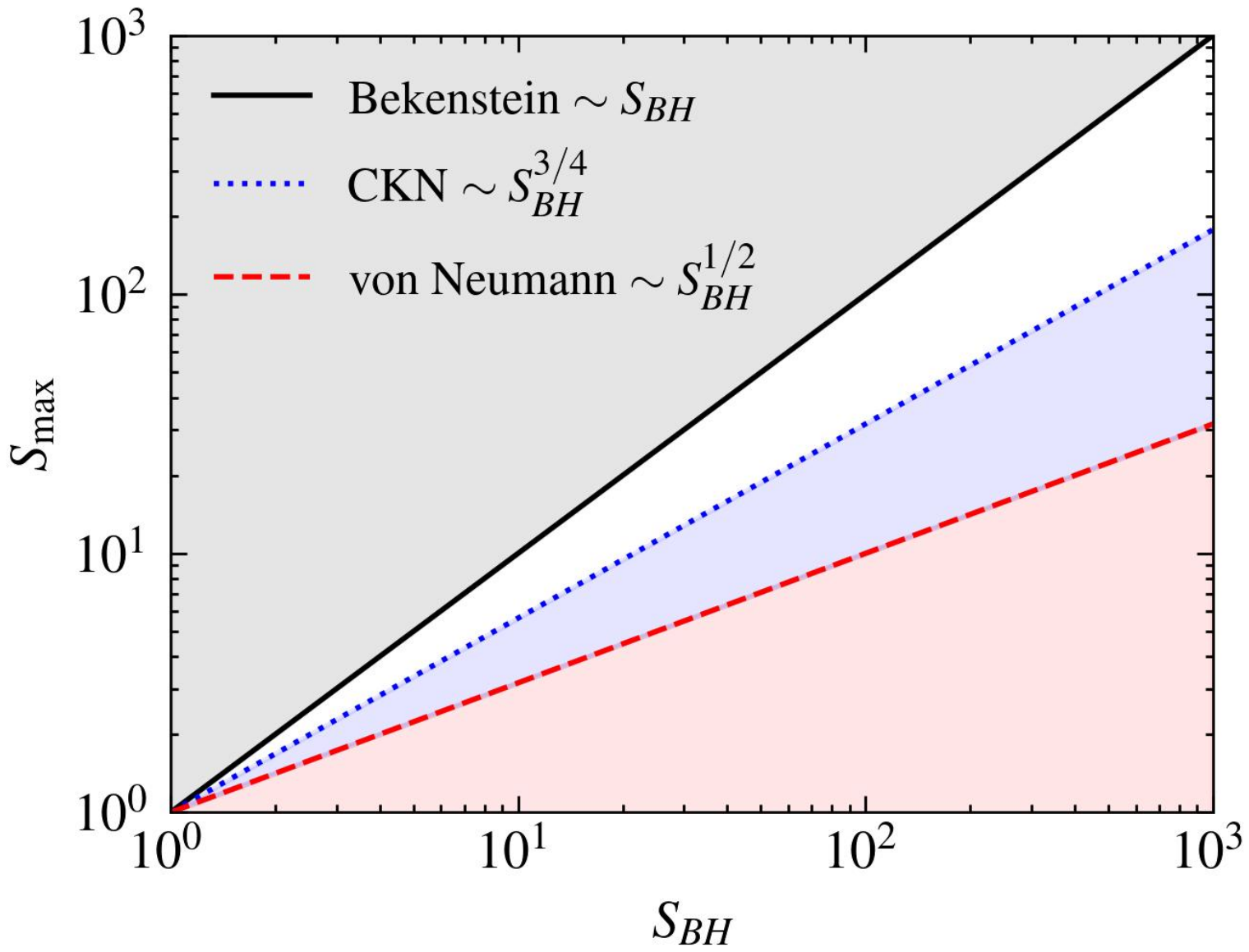}
	\caption{Scaling of entropy bounds against the black hole entropy. Grey: violation of all entropy bounds. White: The CKN and von Neumann bounds are violated. Blue: The von Neumann bound violation. Red: All bounds are satisfied.\label{fig:EntropyBoundScaling}}
\end{figure}

\noindent To summarise:
\begin{itemize}
	\item[1.] The Bekenstein-Hawking Bound
	\begin{itemize}
		\item[$\bullet$] $S{_\text{system (max)}}\simeq S_{\text{Black hole}}$.
		\item[$\bullet$] Here $S{_\text{system (max)}}$ is the maximum possible entropy for any system.
	\end{itemize}
	\item[2.] The CKN Bound
	\begin{itemize}
		\item[$\bullet$] $S{_\text{(max)}}\simeq S_{\text{Black hole}}^{3/4}$.
		\item[$\bullet$] Here, $S{_\text{(max)}}$ is the maximum possible entropy for any system whose entropy is extensive.
	\end{itemize}
	\item[3.] The von Neumann Bound
	\begin{itemize}
		\item[$\bullet$] $S{_\text{inside (max)}}\simeq S_{\text{Black hole}}^{1/2}$.
		\item[$\bullet$] Here $S{_\text{inside (max)}}$ is the maximum possible entropy for any system with boundary ($r_A$) within the IR cut-off such that the total system's entropy is defined to be zero.
	\end{itemize}
\end{itemize}

\textit{Is there any difference in the energy bound}?\\

For the Bekenstein bound, the relation is for an individual system. That is, for a system with energy $E$ and size $R$, the entropy of the system must be less than $\sim ER$. Thus, the maximum energy will be that of a black hole of the same size. In CKN and von Neumann's construction, the energy of the system is taken to be lower than the energy of a black hole of the same size. This immediately does not come from the generalised second law as Bekenstein bound. Thus, unlike the entropy bound, the energy bound in all construction remains the same. Only the definition of entropy changes the bound of entropy. This difference will be more profound once we consider the causal entropy bound. While it would be worth investigating the implications in particle physics, as CKN and CK \cite{cohen2021gravitational} did, our current focus is on exploring only the cosmological aspects.

\section{Dynamic Boundary and Causality}
Although we have utilised the Hubble function, we have not considered any specific real or phenomenological cosmological models. Assuming the concordance cosmology, the $\Lambda$CDM model, we encounter several horizons other than the Hubble scale, such as the particle horizon, future event horizon, and more. In this context, the apparent horizon coincides with the event horizon only in the final de Sitter epoch. Nevertheless, since the laws of thermodynamics apply to the apparent horizon, the concepts of entropy and other thermodynamic quantities remain valid as they do in de Sitter space. However, the laws of horizon thermodynamics are not strictly valid for particle and future event horizon \cite{PhysRevD.92.024001}.

A pertinent question arises: what happens if $r_A$ continues to grow and eventually reaches $L_{IR}$? According to von Neumann entropy, we would have full access to the system's state, and the entropy would be zero by definition. This differs from the scenario CKN considers, where there should be a maximum entropy limit. In the $\Lambda$CDM cosmology, the final de Sitter state represents the desired equilibrium state of maximum entropy. Key points to consider are:
\begin{itemize}
	\item[1.] The final state of the universe with a positive cosmological constant is an equilibrium state with maximum entropy for $\Lambda$CDM cosmology.
	\item[2.] Given a positive cosmological constant, the final entropy corresponds to the Gibbons-Hawking entropy derived from the partition function of the de Sitter horizon.
	\item[3.] Due to the positive cosmological constant in $\Lambda$CDM cosmology, there will be regions of the universe beyond this event horizon that a cosmic observer will never see, regardless of how long they wait.
\end{itemize}

Considering an entropy scaling similar to CKN's, it is not strictly applicable to de Sitter space, as it lacks a boundary like AdS space. In contrast, the von Neumann construction remains valid. In \cite{PhysRevLett.116.201101}, Jacobson shows that the UV entropy contribution depends on the area of the boundary (geometry), and the IR contribution depends on the quantum field (matter), provided the matter fields are at least asymptotically conformal. Therefore, the von Neumann description appears appealing when estimating entropy in the context of cosmology. Thus, in the von Neumann description, the apparent horizon will expand until it reaches the point of maximum entropy and achieves de Sitter equilibrium. So, what is the maximum length scale? It depends on the value of the cosmological constant.

Regarding energy bounds, all descriptions agree on a common principle: the energy density within a given volume of space is less than the energy density defined for a black hole of the same size. Thus, the energy density estimates by CKN and others remain identical.

\subsection{Causal Entropy Bound \& Cosmology}

So far, we have not included the notion of causality in any description. In this context, inflationary cosmology and holography are at odds with entropy bounds \cite{PhysRevD.60.103509}. An interesting approach was proposed in \cite{veneziano1999entropy}, later developed into a more concrete one in \cite{PhysRevLett.84.5695}, using causally connected length scales. The Hubble entropy bound or the causal entropy bound ($S_{CEB}$) proposed in  \cite{veneziano1999entropy} follows a geometric mean of Bekenstein bound and holographic bound such that
\begin{equation}
S_{CEB}=R^3H/\ell_p^2\sim\sqrt{\rho V^2}\sim\sqrt{EV}
\end{equation}
The final piece to the above argument was illustrated in \cite{PhysRevLett.84.5695}, where they identified a causal length scale dubbed $R_{CC}$. The goal was to estimate the size of the black hole that could fit in a given region without falling apart in a cosmological background. This length scale will limit the causally connected regions such that no perturbation beyond this scale can form black holes. Following the Hamiltonian formulation in \cite{BRUSTEIN1998277}, one can identify this length scale as \cite{PhysRevLett.84.5695},
\begin{equation}
R_{CC}=\left\{\text{max}[\dot{H}+2H^2+k/a^2,-\dot{H}+k/a^2]\right\}^{-1/2}
\end{equation} 
Then, using the Einstein field equation, one can show that, 
\begin{equation}
R_{CC}=4\pi G\left\{\text{max}\left[\frac{\rho}{3}-p,\rho+p\right]\right\}^{-1/2}
\end{equation}
Depending on the gravity theory, the form will be modified accordingly. The choice and origin of perturbations can also affect this length scale. However, it is a self-consistent approach where $R_{CC}$ is well defined, and the entropy bound is covariant, provided the weak energy condition is satisfied. 

It is trivial to show that the above bound reproduces the Friedmann equations, as we have already employed the Einstein equation. Nevertheless, for a de Sitter universe, we will have $\dot{H}+2H^2+k/a^2>-\dot{H}+k/a^2$, which is true. For matter and radiation-dominated universe, $\dot{H}+2H^2+k/a^2<-\dot{H}+k/a^2$ holds. Thus, the $R_{CC}$ gets redefined accordingly.

Let us now consider an effective situation with multiple fluids with an effective density $\rho_{\text{eff}}$ and an effective equation of state $w_{\text{eff}}$. Then,
\begin{equation}
R_{CC}=4\pi G\left\{\text{max}\left[\left(\frac{1-3w_{\text{eff}}}{3}\right)\rho_{\text{eff}},(1+w_{\text{eff}})\rho_{\text{eff}}\right]\right\}^{-1/2}.
\end{equation}
We have 
\begin{align}
\left(\frac{1-3w_{\text{eff}}}{3}\right)\rho_{\text{eff}}>(1+w_{\text{eff}})\rho_{\text{eff}}\text{, for } w_{\text{eff}}>-\frac{1}{3}\\
\left(\frac{1-3w_{\text{eff}}}{3}\right)\rho_{\text{eff}}<(1+w_{\text{eff}})\rho_{\text{eff}}\text{, for } w_{\text{eff}}<-\frac{1}{3}
\end{align}
Thus, it differentiates between a decelerating and an accelerating universe. Therefore, for an accelerating universe, such as the current observed one, the causal length scale is $\dot{H}+2H^2+k/a^2$, with the $\rho_{\text{eff}}$ as the effective energy density. Here, irrespective of the length scale, the quantity of interest is the total energy density and not a single fluid's energy density. This is important once we focus on the holographic dark energy models.

\subsection{Implications of Causal Entropy Bounds}

Unlike CKN or von Neumann bounds, the causal entropy bound (CEB) is not built on any explicit assumption of the system. This is done by counting the number of stable black holes that can be formed inside the system. Thus, CEB is the bound for which the respective entropy is covariant. Let us see what happens when we conjecture that each entropy bound at saturation should be covariant.

Let us start with CKN bound, which says that $S{_\text{(max)}}\simeq S_{\text{Black hole}}^{3/4}$. When we demand this to be the same as the causal entropy bound, we get,
\begin{align}
S_{CEB}=S{_\text{(max)}}\simeq S_{\text{Black hole}}^{3/4}\\
\implies V\sqrt{\rho}\simeq R^{3/2}\implies\rho\sim\frac{1}{R^3}
\end{align}
When we demand such equality, we recover the density to scale like ordinary matter. CKN identified a scaling $1/R^2$ because they bounded it by the energy of a black hole, which is different from the CKN entropy bound. To recover the energy bound from the CKN entropy bound, one must assume a relation between energy and entropy, as $(EL)^{3/4}=S$, which depends on the extensive scaling of entropy. Here, when we do not assume any such relation and assume the validity of CEB at saturation, we recover the ordinary matter scaling. 

Similarly, one can repeat the same with other entropy bounds. With the von Neumann bound, we get
\begin{equation}
\rho\sim\frac{1}{R^4}.
\end{equation}
It is interesting to note that this resembles radiation-like scaling. And finally, with the Bekenstein bound, we get
\begin{equation}
\rho\sim \frac{1}{R^2}
\end{equation}
which is the same as the general energy bound.

Thus, to summarise, for a fixed energy system, if the entropy must be a covariant quantity, the density scales like non-relativistic matter with CKN bound, relativistic matter with von Neumann and for Bekenstein bound, the density scales like that of a black hole. This proves that, in the context of cosmology, under covariant entropy bound, different entropy bounds in the literature could correspond to different cosmic fluids/epochs. Thus, each entropy bound reflects different realities rather than being more stringent and distinct.

This observation clarifies what was confusing in the CKN's original proposal. When CKN proposed that $\rho\lesssim 1/L^2$, they were comparing the ordinary energy with that of a black hole. However, when they suggested an entropy bound in the second step, they related extensive entropy with non-extensive black hole entropy through the density scaling. 

In the next part, we explore the origin and construction of holographic dark energy, which claims to have its roots in the CKN entropy bound.

\section{Holographic Dark Energy -- Connections with Thermodynamics}
Holographic dark energy (HDE) is a strong contender for explaining the late-time accelerated expansion. It's been studied extensively, with various attempts to refine and broaden its views \cite{WANG20171}. Initially, the aim was to demonstrate a link between UV and IR phenomena to explain the smallness of the observed cosmological constant. These attempts led to the hypothesis that dark energy might be tied to the horizon entropy, following the area law. However, it's become evident that most approaches to HDE fail to address the original cosmological constant problem.

Most constructions of HDEs start with the expression,
\begin{equation}
\rho_{\Lambda}=3c^2M_p^2L^{-2}_{IR}\propto L^{-2}_{IR}.
\label{eq:SHDE}
\end{equation}
Here, $ c $ is usually a constant, though some models introduce time dependence, $M_p=1/8\pi G$ handles the dimensions, and $L_{IR} $ is the IR cut-off. The standard HDE approach inserts this dark energy density into either standard or modified Friedmann equations to study cosmic evolution. Once we have the Hubble function from this Friedmann equation, we can use various datasets to constrain the models.

Taking a step back from equation (\ref{eq:SHDE}), we have,
\begin{equation}
\rho_{\Lambda}\propto {S}/{L^{4}_{IR}}.
\end{equation}
So, the previous equation corresponds to cases where $S\propto L^{2}_{IR}$, where $ S $ represents horizon entropy. All these constructions trace back to the CKN energy relation \cite{PhysRevLett.82.4971} (not to the CKN entropy bound).

As explained earlier, the original work by CKN \cite{PhysRevLett.82.4971} resolved only the fine-tuning problem. There, the calculation began with a fundamental question: What is the maximum energy possible within a given space? For an ordinary weakly interacting system, the total energy is $E=\rho V$, where $\rho$ and $V\sim R^3$ represent energy density and volume, respectively. Hence, we have $\rho R^3\sim R$, or $\rho\sim R^{-2}$. This yields the bound $\rho\lesssim S/R^4$, where $S\sim R^2$ is the maximum entropy possible in the given space. We have not considered the CKN entropy bound here, as we do not assume any explicit form for the fluid's entropy. Further, in a universe dominated by dark energy, $\rho\sim\rho_{\Lambda}$, with $R\sim 1/H$, we have $\rho_{\Lambda}\sim H^2$. Given $H$ takes the current value of the Hubble function, $\rho_{\Lambda}$ aligns closely with our measurements, thus no fine-tuning. 

The CKN never explained the origin of the cosmological constant, nor did it conflict with general relativity in any way. Their main point was that if we consider a thermodynamic limit in standard field theory calculations, we must account for the UV-IR see-saw \cite{PhysRevD.107.126016}. The only solution to the fine-tuning problem is to get this insight from the CKN bound and implement it in the amplitude estimation of QFT. As emphasised in \cite{PhysRevD.107.126016} and \cite{PhysRevLett.85.1610}, the vacuum energy's smallness implies the universe's immense size. The Hubble scale's choice makes it look ambiguous as they would get an expression similar to the Friedmann equation. This is because the CKN energy relation is consistent with the first law.

In \cite{Moradpour2018}, the authors proposed an alternative to $\rho_{\Lambda}\sim {S}/{L^{4}_{IR}}$ using the first law as,
\begin{equation}
\rho_{\Lambda}\sim TdS/dV.
\label{eq:MoradHDE}
\end{equation}
where $T$ is the horizon temperature. Interestingly, Eq. (\ref{eq:MoradHDE}) also yields similar results and even provides analytical solutions where Eq. (\ref{eq:SHDE}) couldn't. Follow-up studies have revisited these arguments and clarified that both Eq. (\ref{eq:MoradHDE}) and Eq. (\ref{eq:SHDE}) are consistent with the first law itself \cite{Moradpour2024}. Thus, the relationship between HDE and the laws of thermodynamics is undeniable and has been explored from various perspectives \cite{Luongo2017}. 

Now the question is, can we plug this expression back into the Friedmann equation as such? In a previous work, we show that it might be inconsistent to plug this back \cite{Manoharan2023}. The inconsistencies in the entropic cosmology approaches were further demonstrated in \cite{GOHAR2024138781}.

The primary objective of any dark energy model is not merely to account for the late-time acceleration of the universe. While introducing a simple constant might seem sufficient, there are countless mathematical and phenomenological ways to incorporate such a constant. Nevertheless, this approach does not resolve the cosmological constant problem as the CKN. In the following discussion, we will highlight that HDE models, which can account for late-time acceleration, inherently include an integration constant whose value cannot be determined by standard QFT. Consequently, HDE provides, at best, a dynamic solution that closely resembles the $w$CDM model in general scenarios. 
\begin{quote}
	\textit{It is crucial to differentiate between the existence of a constant and the reason for its small value. While CKN addresses the latter, HDE offers only a phenomenological explanation for the former.}
\end{quote}

\subsection{What was wrong with the Hubble cut-off?}

The foundational steps towards the concept of HDE were introduced in \cite{HSU200413} and \cite{LI20041}. These works critiqued the CKN relation's ability to address the fine-tuning problem, arguing that the equation of state was incorrect and that the Hubble scale could not serve as an appropriate infrared cut-off. However, we will demonstrate that this argument needs to be more logical.

In \cite{PhysRevLett.82.4971}, CKN only reached the following conclusions;
\begin{itemize}
	\item The total energy density must scale as $L_{IR}^{-2}$.
	\item Assuming a dark energy-dominated universe, the total energy density should be close to the density of the cosmological constant or any entity playing the role of dark energy.
	\item With the present value of the Hubble function as $1/L_{IR}$, this density aligns closely with the observed dark energy density.
\end{itemize}

However, these assumptions already presume the existence of dark energy, which inherently possesses an equation of state $<-1/3$ (close to $-1$ for anything resembling a constant at present). In other words, CKN was not proposing any dark energy; they were demonstrating its littleness.

The major flaw in \cite{HSU200413, LI20041} arises from circular reasoning. When we substitute the dark energy density of the form $\rho_{\Lambda}\sim H^2$ back into the standard Friedmann equation, we obtain $H^2\sim \rho_m+cH^2$, which implies $\rho_m\sim H^2$. Then, \cite{HSU200413, LI20041} claims that, since the equation of state of $\rho_m$ is zero, $\rho_{\Lambda}$ also has the same equation of state. This argument lacks physical significance and makes no sense.

For instance, if the universe were matter-dominated, CKN would have stated $\rho_m\sim H^2$ from the outset due to its consistency with the first law. In fact, within the framework of the Friedmann universe, all densities scale like $H^2$. One must then know how the specific fluid scales to understand the dynamics of $H^2$.

Let's illustrate this differently. Imagine we were in the matter-dominated era with radiation as the second dominant component, and we knew about the CKN relation. The then-present Hubble function could provide us with the matter density. But what would be the equation of state of matter? Say we have no idea about the matter scaling but know that radiation scales like $a^{-4}$, where $a$ is the scale factor. Based on \cite{HSU200413, LI20041} arguments, one might infer that matter has the equation of state of radiation, as both scales like $H^2$. Hence, claiming that the Hubble scale cut-off gives the wrong equation of state needs to be corrected. 

To summarise, there was nothing inherently wrong with the CKN relation, nor was there an error in estimating energy density using CKN. However, CKN never asserted plugging it back into the Friedmann equation to define a dynamical dark energy.

On the other hand, the dynamic vacuum proposed by \cite{SHAPIRO2009105} hints towards an excellent resolution for the coincidence problem with an integration constant appearing in their construction. Instead of focusing on the running term in \cite{SHAPIRO2009105}, the insight from CKN was to understand that the integration constant should be free of fine-tuning. The additional running nature provided by the RG flow with the Hubble scale may resolve the coincidence. The running nature is anticipated to be a minor corrective term \cite{10.1093/mnras/stab3773}, unlike in HDE, where the dynamic term is the predominant factor.

Thus, it is incorrect to consider HDE a first-order approximation of the running vacuum. The varying nature of the running vacuum arises from the genuine dynamics of the Hubble function as a scale-dependent decay, which can be identified by examining the free parameters. In the CKN-inspired HDE model, the free parameter $c^2$ is close to unity but nearly zero in the running vacuum model. This indicates that the $H^2$ term in the running vacuum model is an RG flow correction, not the primary contributor to dark energy.

An interesting extension of these ideas is the provision of local antigravity sources within FLRW spacetime. Widely known as Swizz cheese models, they can address late-time acceleration without dark energy, but using local sources \cite{PhysRevD.97.123542}. They can be associated with an RG flow associated with Newton's constant and the RG flow of CC \cite{PhysRevD.105.083532}. In \cite{Anagnostopoulos_2019, bonanno2024renormalizationgroupimprovedgravitationalaction}, the authors illustrate how lower IR scales can address cosmic acceleration within asymptotically safe gravity theories. The ability of these models to address ongoing tensions in cosmology is also commendable \cite{Zarikas_2024}.

\subsection{What about particle and future event horizons?} 

The quest to construct a sensible HDE now involves finding the proper IR cut-off. In addition to the Hubble horizon, conventional $\Lambda$CDM models include particle and future event horizons. Although the question of whether it's appropriate to use them to define HDE is set aside, according to \cite{LI20041}, the particle horizon cannot generate the correct equation of state. In contrast, the future event horizon is capable of doing so. Let's explore how these conclusions are derived.

To begin, the particle ($R_p$) and future ($R_f$) event horizons are defined as follows:
\begin{align}
R_p = a\int_{0}^{a}\frac{da'}{a'^2H(a')}\text{ and }
R_f = a\int_{a}^{\infty}\frac{da'}{a'^2H(a')}
\end{align}
where $a$ is the scale factor. A straightforward approach assumes a dark energy-dominated universe such that $H^2\sim \rho_{\Lambda}$. Consequently, we obtain a dark energy density, $\rho_{\Lambda}={3\alpha^2}a^{-2\left(1\pm\frac{1}{c}\right)}/({8\pi G})$.
Here, $+$ and $-$ correspond to $R_p$ and $R_f$, respectively. Notably, a positive value of $c$ favours $R_f$ as the preferred choice. However, a crucial point overlooked is the existence of the constant $\alpha$. This new constant has absorbed the $c$ previously present in the expression and appears as an integration constant. Thus, regardless of the dynamic nature offered, the present value relies on the initial conditions, shifting the fine-tuning problem to an initial value problem. With $c=1$ in the case of $R_f$, we recover the cosmological constant, and $\alpha$ would become unity by definition for a dark energy-dominated universe. Does it answer why this is small? NO!

Even if we acknowledge this characteristic of HDE construction, further aspects must be considered. Assuming the future event horizon entails a logical flaw because it presupposes accelerated expansion. Therefore, the analysis introduces circular reasoning by assuming acceleration to derive acceleration, diminishing the future event horizon as a compelling candidate.

\section{Why is only Dark Energy considered holographic in HDE?}
Considering the CKN relation, there are notable distinctions in the formulation of HDE. Unlike the CKN relation, which does not explicitly differentiate between energy components, HDE posits that only dark energy is holographic. Therefore, the dark energy term must originate from the geometric aspect when deriving the Friedmann equations of HDE from a suitable action. Indeed, possible ways need to be further verified \cite{PhysRevD.88.023503}.

The construction of HDE using the future event horizon highlights that the density of dark energy also relies on the density of matter throughout all epochs. This implies a potentially non-trivial coupling in the action. However, in cosmological studies, HDE consistently assumes the standard Friedmann equations, which could be considered a limitation. Although attempts have been made to construct action for HDE models \cite{li2012newmodelholographicdark}, how they differ from modified gravity theories is not clear. As in most cases, there can be one to one correspondence \cite{DAGOSTINO2024138987}. 

Another significant observation is that the equation of state of dark energy is dynamic, and its precise characteristics depend on the cosmic components under consideration. We will explore these aspects further when examining alternative infrared cut-offs involving derivatives of the Hubble function.

\subsection{Infrared Cut-off with Derivatives of the Hubble Function}

A compelling and more coherent approach involves considering an infrared cut-off utilising derivatives of the Hubble function \cite{doi:10.1142/S0218271812500915}. This method ensures causality and circumvents the circular reasoning associated with the future event horizon. Conceptually, one can liken this approach to incorporating $f(R)$ terms in an action \cite{Rezaei2022}. Furthermore, this approach finds motivation from the development of running vacuum models \cite{SHAPIRO2009105}. Once again, the fundamental questions emerge: why does this approach succeed in accounting for late time acceleration, and does it effectively address the fine-tuning problem?

The key reason this approach is compelling stems from an integration constant. The dynamic behaviour, or equation of state, of dark energy, is contingent upon the values of all other independent parameters. Although these parameters may have theoretical constraints, their precise values must conform to observational constraints. Given that late-time acceleration is empirically established and the equation of state of dark energy approximates $\sim -1$ (at least its present value), the parameters within the model must adhere to these observational constraints. Hence, the model's success in explaining late-time acceleration hinges on identifying parameter values consistent with observational data. 

Several infrared cut-offs involving derivatives of the Hubble function are explored in the literature. The Ricci scalar, termed Ricci HDE, stands out for its adequate explanation of observations \cite{PhysRevD.79.103509, PhysRevD.79.043511}. Introducing additional flexibility to this cut-off gives rise to the Granda-Oliveros cut-off, offering a broader perspective \cite{GRANDA2008275}. However, this extension can lead to scenarios where the dark energy density becomes negative, which doesn't violate physical principles since the total energy density remains positive by construction \cite{Manoharan2024}. Interestingly, the negative dark energy density in the past might help resolve anomalies like the BAO Lyman-alpha tensions, as suggested recently \cite{Tiwari2024}. Additionally, other extensions involve Gauss-Bonnet terms discussed in the literature \cite{PhysRevD.97.064035}. 

The Granda-Oliveros cut-off, represented by $ L_{GO} = 1/\sqrt{\alpha H^2 + \beta \dot{H}} $, lacks a precise theoretical foundation unlike the Ricci or Gauss-Bonnet cut-offs but extends the range of possibilities in cosmology. When substituted into the CKN relation, it leads to the expression $ \rho_{\Lambda} = 3M_p^2 (\alpha H^2 + \beta \dot{H}) $. This relation, when inserted back into the first Friedmann equation and solved for $ H $ (neglecting radiation and curvature), gives a modified form of the Hubble function as,
\begin{equation}
H = H_0 \sqrt{\frac{2 \Omega_{m0} a^{-3}}{-2\alpha + 3\beta + 2} + \left(1 - \frac{2 \Omega_{m0}}{-2\alpha + 3\beta + 2}\right) a^{\frac{-2(\alpha - 1)}{\beta}}}
\end{equation}
This equation indicates a modified matter density and a dynamic dark energy density. Notably, when $ \alpha = 1 $, the dynamic dark energy behaves akin to a cosmological constant, and for $ \beta = \frac{2}{3} $, it aligns with the standard $ \Lambda$CDM model (excluding radiation). The above equation can be reformulated into an effective $w$CDM model, rendering the HDE construction irrelevant. Moreover, this formulation leads to negative energy densities when radiation is considered \cite{Manoharan2024}.

GO cut-off appears more general than Ricci and Gauss-Bonnet cut-offs. However, the latter two are invariants and naturally emerge in the constructions of general action. In the covariant generalised description, we have the IR cut-off depending on various length scales and invariants \cite{Nojiri2017epjc}. However, what better explains the data needs to be noted. For instance, with Ricci or Gauss-Bonnet HDEs, the correction factors to the Hubble function are factors with derivatives of the Hubble function with a correlated coefficient. If this correlation is expected to be natural and the best-fit case, then we must recover them from the observational data. However, fitting GO-HDE do not recover a Ricci-like correlation between the free parameter \cite{Manoharan2024, 10.1093/mnras/stae2257}. Thus, although there are natural options with invariant scalars, keeping the option open is better for data-driven explorations.

Furthermore, another insight emerges from the second Friedmann equation. Using the non-interacting standard continuity equation, we have,
\begin{equation}
\dot{H}+H^2=\frac{-4\pi G}{3}\left(\rho+3p\right).
\end{equation}
Here, $\rho$ denotes the total density, and $p$ represents the corresponding pressure. This equation can be reformulated using the first Friedmann equation,
\begin{equation}
\frac{3}{8\pi G}\left(H^2+\frac{2}{3}\dot{H}\right)=-w_{\text{eff}}\rho
\end{equation}
Taking $w_{\text{eff}}=-1/\tilde{\alpha}$, we derive,
\begin{equation}
\rho=\frac{3}{8\pi G}\left(\tilde{\alpha}H^2+\frac{2\tilde{\alpha}}{3}\dot{H}\right)
\end{equation}
Assuming $\tilde{\alpha}$ as $\alpha$ and $2\tilde{\alpha}/3$ as $\beta$, this expression resembles the HDE density utilizing $L_{GO}$. While this formulation corresponds to the total density with an effective equation of state $w_{\text{eff}}$, it bears a striking resemblance to the Granda-Oliveros HDE dark energy density.

Examining the dark energy equation of state derived from the Granda-Oliveros HDE model unveils a dynamic character where dark energy acts as the dominant energy component in the respective era. Consequently, during the matter-dominated epoch, the dark energy equation of state approaches zero, while in the radiation-dominated phase, it approximates $ 1/3 $. Transitions between these phases may not always occur smoothly, potentially leading to shifts towards negative energy density. This observation highlights the similarity between the second Friedmann equation and the GOHDE model.

The model eliminates radiation-like behaviour when $\beta=\alpha/2$, and similarly avoids matter-like behaviour with $\beta=2\alpha/3$. This similarity between the Granda-Oliveros HDE model and the total density reinforces its alignment with the first law of thermodynamics. This raises the question of whether this framework can exclusively describe dark energy.

Furthermore, the analysis above assumed that $w_{\text{eff}} = -1/\tilde{\alpha}$, where $ w_{\text{eff}} $ denotes the effective equation of state—a dynamic quantity. Hence, its inherent construction does not consider $ \tilde{\alpha} $ constant. However, when considering only dark energy and matter, $ w_{\text{eff}} $ can be replaced with $ w_{DE} $, which remains constant. Therefore, selecting values for $ \alpha $ and $ \beta $ near or equal to $ 1 $ and $2/3$ provides a coherent description for late-time scenarios, but not when radiation is involved.

As we consider earlier cosmic phases, the equation of state for GOHDE density transitions to $1/3$, signalling behaviour akin to radiation. The radiation component must become supercritical to accommodate this transition, potentially resulting in negative dark energy density. Addressing this challenge might require transitioning from $ \beta = 2\alpha/3 $ to $\beta = \alpha/2 $. However, existing HDE constructions do not naturally accommodate this. When we add a constant along with this HDE, the construction looks identical to the running vacuum, with different values for the free parameters.

The close resemblance of GOHDE and the second Friedmann equation suggests constructing a unified dark sector rather than dark energy alone. This observation is simply due to the CKN relation, which considers the total energy density. The success route for HDE might be identifying the whole dark sector with holographic energy rather than dark energy alone.

When linking the origin of HDE to the CKN energy relation, it is crucial to recognize that we are modelling the total energy density, not just dark energy. This connection remains consistent across different entropy bounds, as the energy bound is the same for both Bekenstein and von Neumann bounds. The key distinction lies in how entropy scales in each case. Therefore, understanding the scaling of dark energy—essentially its equation of state—is vital for modelling it. For example, with its constant density, the cosmological constant differs from black holes, whose density scales inversely with their area. When this area becomes constant, the respective density becomes constant, which resembles the de Sitter solution (the CC). As long as the information about this constant area is missing, we cannot fix the value of CC. Thus, the knowledge about the value of the concordance CC lies in the value of the Hubble function at $z=-1$. If this dark energy is not a constant, as recent DESI results indicate \cite{adame2024desi}, the CC problem will get diluted further. If the dark energy density must scale like a constant, then according to CEB discussed earlier, we have,
\begin{eqnarray}
V\sqrt{\rho}&=S_{\text{(max)}}^n~ \text{, with constant }\rho\\&\implies V\propto S_{\text{(max)}}^n\nonumber\\
&\implies R^3\propto R^{2n}\nonumber\\
&\implies n=3/2
\end{eqnarray}
which violates the Bekenstein bound, when $S_{\text{(max)}}\sim R^2$. In short, a specific entropy scaling and bound can only be definitively assigned to dark energy with deeper insights into its nature.
\section{Summary}
To summarise, the CKN energy relation stems from the first law of horizon thermodynamics, particularly at saturation. This implies that, in theory, the CKN relation for energy density serves as a limit on the total energy density within a region of space rather than directly modelling dark energy. Consequently, the values from different periods align closely with the dominant energy density. From a cosmologist's perspective, the CKN bound is closely tied to the first law of horizon thermodynamics and the first Friedmann equation. We also derived a new entropy bound called the von Neumann entropy bound based on non-extensive scaling. As other entropy bounds also share the same first law, the energy bound remains invariant. When it comes to the entropy scaling of dark energy, we cannot connect it to any particular form. In other words, different entropy bonds will have different consequences. Under the causal entropy bound (CEB), the underlying bound must violate the Bekenstein bound to have a constant density scaling.

Many modified gravity theories and most HDE frameworks use an integration constant to explain late-time acceleration. However, this approach does not address the fundamental cosmological constant problem. The key insight from the CKN bound is the concept of UV-IR mixing, which reduces the effective degrees of freedom in EFT. Thus, as explored in \cite{PhysRevD.94.104052, PhysRevD.107.126016} with a modular structure for quantum space, the CKN bound should be applied in QFT computations to address the original CC problem. While most alternatives to the standard CC and Einstein gravity attempt to explain the coincidence problem, HDE offers a solution by providing a tracking mechanism that naturally evolves into a cosmological constant without requiring arbitrary interactions between fluids.

HDE models, in general, do not preserve the dynamics of other fluids when considering the overall conservation equation. If complete conservation is assumed, the resulting solutions resemble running vacuum models, affecting the behaviour of matter, radiation, etc. The main difference between HDE and running vacuum models lies in how the integration constant is introduced. In running vacuum models, the integration constant appears at the field equation level from matter lagrangian, whereas in HDE, it is incorporated at the level of the Friedmann equations. Thus, HDE models with derivatives of the Hubble function are more practical at the Friedmann equation level, avoiding issues such as causality problems associated with future event horizons. However, the value of the integration constant cannot be predicted, and the action level origin remains obscure.

We did not discuss the possibility of redefining horizon entropy and deriving field equations using Wald's formalism \cite{PhysRevD.48.R3427, PhysRevD.50.846}. A modified entropy framework is theoretically equivalent to a modified gravity theory. For instance, Tsallis entropy can lead to an $f(R)$ gravity model, which simplifies to Einstein gravity in the appropriate limit \cite{DAGOSTINO2024138987}. In this framework, the modified non-extensive parameter causes the late-time acceleration with a modified integration constant that depends on the extra parameter. Similar approaches have been explored using Barrow \cite{PhysRevD.107.023505, PhysRevD.108.103533}, Renyi \cite{Manoharan2023}, and other non-extensive entropy formulations. In most cases, dark energy exhibits a tracking behaviour, following the dominant energy component. Thus, it behaves naturally like a cosmological constant in a dark energy-dominated era and like matter in a matter-dominated era. However, the CC-like behaviour needs a CC in the first place, which comes in as an integration constant whose value is not predicted by the model.

\begin{acknowledgments}
Thanks to David B. Kaplan for clarifying doubts regarding the CKN bound and to Titus K. Mathew for discussing HDE models. Thanks to the referees for pointing out valuable references and comments which improved the presentation of this manuscript. This research was supported by CSIR-NET-JRF/SRF, Government of India, Grant No: 09 / 239(0558) / 2019-EMR-I. 
\end{acknowledgments}

\end{document}